# Thermally-induced magnetic order from glassiness in elemental neodymium

Benjamin Verlhac[1], Lorena Niggli[1], Anders Bergman[2], Umut Kamber[1], Andrey Bagrov[1,2], Diana Iuşan[2], Lars Nordström[2], Mikhail I. Katsnelson[1], Daniel Wegner[1], Olle Eriksson[2,3], Alexander A. Khajetoorians[1,*]

1. Institute for Molecules and Materials, Radboud University, Nijmegen, The Netherlands

2. Department of Physics and Astronomy, Uppsala University, Uppsala, Sweden

3. School of Science and Technology, Örebro University, SE-701 82 Örebro, Sweden

*corresponding author: a.khajetoorians@science.ru.nl

**Temperature in thermodynamics is synonymous with disorder, and responsible for ultimately destroying ordered phases. Here, we show an unusual magnetic transition where, with increasing the temperature of elemental neodymium, long-range multi-Q magnetic order emerges from a self-induced spin glass. Using temperature-dependent spin-polarized scanning tunneling microscopy, we characterize the local Q order in the spin-Q glass phase and quantify the emergence of long-range multi-Q order with increasing temperature. We develop two distinct analysis tools, which enable the quantification of the glass transition temperature, based on measured spatially-dependent magnetization. We compare these observations with atomic spin dynamics simulations, which reproduce the qualitative observation of a phase transition from a low-temperature spin glass phase to an intermediate ordered multi-Q phase. These simulations trace the origin of the unexpected high temperature order in weakened frustration driven by temperature-dependent sublattice correlations. These findings constitute an example of order from disorder and provide a rich platform to study magnetization dynamics in a self-induced spin glass.**

Ordered phases tend toward disorder with increasing temperature, resulting from entropy. There are a few exceptions to such intuition, as exemplified by the Rochelle Salt, the first ferroelectric discovered. In this material, there are two relevant Curie temperatures, where for both $T < T_{C1}$ and $T > T_{C2}$, the material exhibits a disordered paraelectric phase, but oddly an ordered ferroelectric phase stabilizes for $T_{C1} < T < T_{C2}$ [1]. For magnetic systems, no such example is known to date. A study on $Y_2Ni_7$ was initially interpreted as thermally induced spontaneous magnetization [2], but this was later found to be flawed by sample impurities [3]. Theoretically, there are mechanisms predicting the appearance of ferromagnetic order at increasing temperature [4]. For example, in itinerant electron systems, ferromagnetic order may arise if a peak in the density of states shifts such that the Stoner criterion is satisfied at finite temperature, but not at $T = 0$. Another heavily discussed mechanism to create magnetic order from an increase in temperature is known as order-from-disorder and is related to a macroscopic degeneracy of the ground state that is broken by finite-temperature contributions of spin excitations to the free energy [5,6]. These proposed mechanisms provide counterexamples to basic thermodynamic intuition, where increasing temperature, and thus increasing entropy, should produce more disorder.

Spin glasses are a special class of magnets exhibiting a critical temperature, while still lacking long-range order below this temperature. They are often considered to be disordered magnets. Spin glasses are distinguished from both equilibrium and conventional nonequilibrium systems by exhibiting aging, that is, the presence of a peculiar dynamics characterized by a very broad distribution of relaxation times [7,8]. The latter distinguishes a spin glass state from other frustrated magnets, like spin ices and spin liquids [9,10]. Traditionally, the origin of spin glass behavior resides in the combination of competing magnetic interactions and significant disorder [11-14]. Recently, a new type of spin glass, the so-called self-induced spin glass [15-17], was discovered to be the magnetic ground state of elemental crystalline neodymium [18]. Unlike conventional spin glasses, the self-induced spin glass state is caused solely by competing interactions derived from the lattice structure itself, i.e. in the absence of disorder. In the rare-earth element Nd, these competing exchange interactions lead to a multiplicity of low-energy states defined by a reciprocal lattice vector, or magnetic wavevector, Q. These Q states exhibit strong local Q-order, but with varying periodicity and a complex Q-dependent aging behavior. In contrast to conventional spin glasses, different experimental methods have identified numerous disputed phase transitions below the ordering temperature ($T_N$) of Nd [19-24]. Counterintuitively, these phases have been interpreted as long-range magnetic order, which is at odds with the observation of a low temperature self-induced spin glass[18].

Here, we show that the self-induced spin glass state of elemental neodymium exhibits long-range ordered multi-Q phases at temperatures above the apparent glass transition temperature. Using temperature-dependent spin-polarized scanning tunneling microscopy (SP-STM), we imaged the emergence of long-range order from the spin-Q glass state, as a function of increasing temperature. Using atomic-scale imaging, we quantified the various short-range and long-range Q states. In order to analyze the emergence of long-range order, we developed two analytical tools that quantify the observed phase transition temperature, based on statistical analysis of the spatially dependent magnetization images. Using atomic spin dynamics simulations (ASD), we qualitatively reproduce the two observed magnetic phases below the Néel temperature, and relate this complex behavior to the competing interactions driven by the two sublattices of the double hexagonal close-packed (dhcp) crystal structure.

The spin-Q glass state of Nd is characterized by the presence of so-called Q-pockets [17,18], which are each defined by a distribution of favorable Q states. The favorability of multiple Q states is driven by the competing exchange interactions, linked to the dhcp structure of Nd (Fig. 1a). The presence of Q-pockets leads to self-induced spin glass behavior and can be measured by magnetization images taken on the Nd(0001) surface using SP-STM (Fig. 1b) (see Methods section). The spin-Q glass state can be distinguished by small spatial regions, on a length scale in the order of 10-100 nm (as schematically depicted in Fig. 1c), with well-defined local periodicity (Fig. 1e-g). However, the multiplicity of Q states leads to a varying spectral distribution of Q states, depending on the local region, and varies spatially across the surface (Fig. 1c). The local periodicity can be extracted using Fast Fourier Transforms (FFT) of the magnetization images, or so-called Q-space images, which reveal the periodicities of the local magnetic order in a given region (Fig. 1h-j). By imaging regions with a statistically significant amount of spins, Q-space imaging reveals the relevant Q-pockets at a given temperature (Fig. 1d). For sufficiently low defect density (here ~0.0025 ML where 1 ML corresponds to a monolayer of Nd(0001)), all relevant Q-pockets ($Q_A$ = 1.1-1.5 nm$^{-1}$, $Q_B$ = 1.75-2.9 nm$^{-1}$ and $Q_C$ = 4.1-5.5 nm$^{-1}$) are visible at $T$ = 5.1 K, including the smallest Q-pockets which were previously only observed at lower temperature [18]. We note that the reciprocal atomic lattice would have corresponding spots in a given Q-space image at magnitudes of 17.18 nm$^{-1}$, which is too large to be illustrated in all given Q-space maps.

In order to quantify the phase diagram of the spin-Q glass state, we performed temperature-dependent magnetization imaging on a given area, below the reported Néel temperature ($T_N$ = 19.9 K) [25]. In Fig. 2, we illustrate the same imaged area from Fig. 1b, (200 x 200 nm$^2$), first measured at $T$ = 11 K (Fig. 2b), and reimaged at $T$ = 5.1 K (Fig. 2a). In comparison to the low-temperature spin-Q glass phase, the 11 K image exhibits large-scale domains, separated by well-defined domain walls, with long-range and well-defined multi-Q order. At first glance, all low Q states have disappeared. Using the Q-space image (Fig. 2d), we quantify the values of the multi-Q states, which we describe as diamond-like and a stripe-like patterns in the magnetization image, respectively. Close-up views of these patterns and their respective multi-Q states are shown in Fig. 2e-f. We emphasize that from the resolution of the FFT (Fig. 2d), the accuracy of the q value is about ±0.07 nm$^{-1}$ while the angular precision is about ±1.4°. Within this accuracy, we find that the multi-Q state of the diamond-like patterns (Fig. 2g) can be described as a combination of the Q-vectors $q_1^D$ and $q_2^D$, both with magnitude 2.64 nm$^{-1}$ and oriented along the $[\bar{1}2\bar{1}0]$ and $[\bar{1}\bar{1}20]$ directions, respectively. The multi-Q state of the stripe-like patterns (Fig. 2h) can be expressed as a combination of three Q-vectors: $q_1^S$ and $q_2^S$ with magnitudes 2.41 nm$^{-1}$ and 2.69 nm$^{-1}$ and oriented near the $[2\bar{1}\bar{1}0]$ and $[\bar{1}\bar{1}20]$ directions, respectively, as well as $q_3^S$ with magnitude 5.12 nm$^{-1}$ and oriented close to the $[\bar{1}2\bar{1}0]$ direction. Different from those of the diamond phase, these Q-vectors are not exactly aligned with the crystallographic directions, and angles between the Q-vectors of the stripe phase vary between 120° and 115°. We note that the multi-Q states of the two observed phases present similar vectors, which leads to an apparent blurring/splitting of certain spots when averaging over all domains (see Fig. S4). After cooling back to $T$ = 5.1 K (Fig. 2a), the long-range ordered multi-Q state disappears and the spin-Q glass state reappears with an aged Q-state distribution that, when comparing to the pre-annealed sample, exhibits aging. By further temperature cycling (see Supplementary Section S2), we observed the same ordered multi-Q state at higher temperature, without any indications of aging, whereas we always observed aging of the spin-Q glass states at the lowest temperature. This observation clearly indicates that there is a glass transition temperature ($T_G$), at a temperature below $T_N$, which separates the glassy state from a well-defined ordered state.

In order to quantify the glass transition temperature and the multi-Q phase, we measured magnetization images of the same spatial area at several temperatures. The relevant Q-space images for $T$ = 5.1 K, 6.6 K, 7.5 K, 8.9 K and 11 K are illustrated in Fig. 3a-e. The corresponding magnetization images can be found in Fig. S5. For $T$ < 8 K, the observed Q-pockets are extremely sensitive to the given sample temperature. There is a smooth trend with increasing temperature toward fewer and sharper Q-pockets. The lowest and highest magnitude Q-spots fade with increasing temperature, eventually leading to the well-defined multi-Q state. The observation of a multi-Q state with well-defined domains can be seen from $T$ = 8.3 K and remains robust up to the highest measured temperatures ($T$ = 15.6 K). We note that above 15.6 K, we observed an increase of surface contamination along with diffusion of defects, limiting the temperature range ($T$ < 15.6 K) at which we could image and properly compare.

In order to precisely extract $T_G$, we developed two distinct analytical methods which utilize the spatially dependent magnetization images as a function of temperature and derive a value that is related to the evolution of long-range order. Both methods utilize the fact that the spin-Q glass exhibits strong local order, with well-defined Q vectors in small spatial regions [18]. The distinguishing feature between the ordered and glass phase is that the local order is typically unique in the spin-Q glass phase, whereas it represents a global character in the multi-Q phase. With the first method (see Supplementary Sections S4), we sampled a number of 40 of random 22 by 22 nm$^2$ areas, for each temperature, and extracted the corresponding Q-space images. Based on this, we computed the divergence of each image with the average of all previous images. We then performed this analysis 30 times, for one temperature, and computed the average of the divergence (Q-state divergence, $\mathcal{D}_Q$), and the standard deviation (error bars). The resulting temperature dependence of $\mathcal{D}_Q$ extracted from analyzing numerous magnetization images is plotted in Fig. 3f. This method clearly captures the sensitive changes in the spin-Q glass state with increasing temperature, and $\mathcal{D}_Q(T)$ exhibits a clear plateau above a critical temperature, which correlates with the appearance of the multi-Q phase. Using least squares fitting, we can extract a transition temperature $T_G$ = 8.1±0.3 K. The finite value of the Q-state divergence in the multi-Q phase can be attributed to the appearance of multiple rotational/mirror domains, domain walls, and defects.

The second analytical method to extract $T_G$, is the computation of the complexity ($\mathcal{C}$) in images, which was recently developed based on computing dissimilarities in patterns at different length scales[26]. It was demonstrated that the derivative of this quantity is extremely sensitive to magnetic phase transitions as well as changes in magnetic patterns. We computed the complexity, $\mathcal{C}(T)$, as a function of temperature (see Supplementary Section S5) to analyze the same set of images used in the Q-state divergence analysis in Fig. 3f. Similarly to $\mathcal{D}_Q(T)$, two distinct regimes can be identified in $\mathcal{C}(T)$. Again using linear analysis, we can extract $T_G$ = 7.9±0.2 K. We note that a slightly modified procedure of the complexity analysis leads to $T_G$ = 8.1±0.2 K (see Supplementary Section S5). Both values are in accordance with the $T_G$ value extracted from the divergence analysis. The differences in $\mathcal{C}(T)$ and $\mathcal{D}_Q(T)$ can be attributed to the variations in the method in how both noise and impurities affect the results.

To further understand the physical origin of the spin-Q glass to multi-Q transition, we performed atomistic spin dynamics (ASD) and Monte Carlo (MC) simulations based on exchange parameters obtained from *ab initio* electronic structure calculations (for a review, see Ref. [27]). We simulated the magnetic ground states and thermodynamic properties for a range of temperatures between 1 and 15 K. The specific heat, which was obtained from the derivative of the total energy with respect to temperature [27], is shown in Fig. 4a and exhibits two distinct peaks. Since a peak in the specific heat indicates a magnetic phase transition, our spin simulations can thus identify two different phase transitions with increasing temperature. The low-temperature state was identified as the spin-Q glass state in Ref. [18]. An analysis of the two-time autocorrelation function (Supplementary Section S6), which can be extracted from the simulations, confirms that Nd exhibits the spin-Q glass phase in the whole range of $T$ = 0-4 K. In contrast, for the intermediate temperature range 4 K < $T$ < 11 K, the simulations did not indicate any signs of glassy dynamics. The magnetic order in this range was determined by the static correlation function $S(\mathbf{q})$ (see Methods for computational details) and could be characterized as a traditional multi-Q phase. Above the second phase transition temperature at $T$ = 11.5 K, the system becomes paramagnetic. While the precise values of temperature show small discrepancy with the experimental data, the phase diagram shows strong qualitative agreement with the experimental findings. Namely, there exists below the computed $T_N$ (11.5 K), first an ordered phase driven by broken frustration and ultimately a self-induced spin glass phase.

To explain what drives the spin-Q glass behavior and causes the transition to the multi-Q state we investigated the role of the two different sublattices present in the dhcp structure. In Ref. [18], a significant difference between the magnetic interactions for the different sublattice sites was identified, and here we have investigated the sublattice effect further by studying the temporal and spatial correlations between Nd atoms at different sublattice sites. In Fig. 4b we show static correlation functions obtained by only considering correlations between atoms on the same sublattice, i.e. atoms on cubic (cub) and hexagonal (hex) sites in the system. We observed significant differences between the sublattice correlations, where the cubic sublattice correlations exhibit not only different maxima than the hexagonal sublattice correlations but also a different temperature behavior. From our simulations we observed that the cubic-cubic correlations vanish at the first phase transition temperature while the hexagonal-hexagonal correlations prevail up to the paramagnetic transition. These results can be interpreted as follows: At low temperatures, both sublattice magnetizations have different tendencies toward magnetic order, i.e., they would prefer two different magnetic orderings. It is the competition between these ordering tendencies that effectively results in the spin-Q glass behavior. The effective strength of the exchange interactions on the cubic sublattice is weaker than the interactions on the hexagonal sublattice, which results in the cubic correlations vanishing at a lower temperature, i.e. the first phase transition at $T$ = 4 K. Without the competition from the magnetic atoms on the cubic sublattice, the magnetic order driven by the atoms on the hexagonal sites dominates and the result is then a non-glassy multi-Q state which persist up to the paramagnetic transition.

Our experimental finding and theoretical analysis partly support but also expand the interpretations from neutron diffraction, susceptibility, heat capacity, thermal expansion and other measurements [19-24]. Firstly, we confirm that there is an ordered multi-Q phase above $T_G \approx 8.1$ K (which has been referred to as $T_3$ in the literature), driven by coupling of the hexagonal site spins. However, while neutron diffraction interpreted the multi-Q phase as a relatively simple "double-$q$" structure [28], our data clearly shows that various domains with in total five different $q$ vectors exist. The fact that these (and their higher harmonic) vectors are very close to each other can easily obscure their presence in reciprocal space, when averaging over many domains (as exemplified in Fig. S4e). This may explain the simplified interpretation of the neutron diffraction data, and underscores the need for a spatially resolved technique to unravel the complex magnetic structure of Nd. Secondly, we confirm that the phase transition is driven by the onset of coupling of the cubic site spins. This also illustrates that the surface magnetization pattern, as probed by SP-STM, is representative of the bulk. However, while the phase below 8 K was interpreted as a "triple-$q$" magnetic structure [19,22], we identified it as the glass transition point, below which no long-range order can be found anymore. We note that we did not observe significant changes of the magnetic structure around 6 K in comparison to higher temperature, where previous studies identified another phase transition near 6 K that was interpreted as a "quadrupole-$q$" structure [20-22,24]. We merely observed gradual changes in the distribution of and within the various Q-pockets. The most important effect is the gradual appearance with decreasing temperature of $Q_A$-pockets as well as pockets at angles far off the crystallographic axes. While the former was not observed in neutron diffraction, the latter could be easily misinterpreted as stemming from a fourth vector. We would like to emphasize that the spin-Q glass phase is very sensitive to even small defect concentrations. In a previous study with a typical surface defect concentration of ~0.010 ML [18], we found that the $Q_A$-pockets had already "melted" away at 4.2 K. In the present data, with four times less defect concentration, the $Q_A$-pockets are visible up to $T_G$. The defect-induced pinning of $Q$ vectors, leading to "less glassiness", may also account for the deviating observations found in neutron diffraction.

In conclusion, we have demonstrated the emergence of an ordered phase when the spin-Q glass observed for elemental neodymium is heated above the glass transition temperature, which is a first of its kind in magnetic materials. The origin of this unexpected ordering can be traced to quenching of one set of sublattice spin correlations, which is the essential ingredient for driving the spin-Q glass phase. In magnetic systems, it is known that frustration induced by competing interactions can lead to both complicated (e.g., noncollinear) magnetic ordering and to spin-glassiness. We demonstrate the first example of a phase transition between these states and show that the combination of frustration and temperature can lead to very counterintuitive behavior, with the emergence of regular and ordered magnetic patterns at relatively higher temperature. The example of Nd provides a unique platform to investigate complex magnetization dynamics in a system with strong short-range spin correlations which can be tuned by temperature. This may provide a platform to study dynamical behavior present in glass systems, such as dynamic heterogeneity. This gives a solid experimental and material-specific theoretical background for the concept of order from disorder which is claimed to be important far beyond physics [17,29].

**Methods**

Nd islands were epitaxially grown using the Stranski-Krastanov method on a cleaned W(110) (see Supplementary Section S1). The thicknesses of the islands were around 100 ML and represent bulk-like structural, electronic and magnetic properties [18]. The experimental study was performed in a commercial Createc LT-STM/AFM, operating at a base temperature of 5 K. The temperature-dependent study was done by means of a Zener diode attached to the SPM head, allowing to reach stable conditions between 5 K and 20 K. To get spin contrast we used both Cr bulk tips and Nd-coated W tips. They showed similar results, and all data shown in the main text were acquired using Cr tips, while the aging study (see Supplementary Section S2) was done using Nd-coated W tips. The magnetization images were produced by subtracting majority and minority SP-STM images [18], and a median filter (1x6 pixels) was applied before calculating the FFT. No further data processing was applied to the Q-space images shown. The image data processing was performed using MATLAB.

The simulations for the static correlation functions were done on a slab of 196 by 196 atoms with a thickness of 16 dhcp layers. The system was described using the same spin model as in Ref. [18], i.e., with a Heisenberg Hamiltonian using scalar Heisenberg exchange interactions calculated from DFT. The simulation protocol consists of first equilibrating the system using $10^5$ Monte Carlo sweeps and then performing ASD simulations in order to obtain a thermal average of the static correlation function $S(q)$. The specific heat data was obtained by taking the numerical derivative with respect to the temperature of the total energy obtained from the same simulations used to calculate $S(q)$. The simulations were done using the UppASD software [30], and all algorithms used are given in full detail in Ref. [27].

**Acknowledgments**

This project has received funding from the European Research Council (ERC) under the European Union's Horizon 2020 research and innovation programme (grant agreement No 818399). A.A.K. acknowledges the NWO-VIDI project "Manipulating the interplay between superconductivity and chiral magnetism at the single-atom level" with project number 680-47-534. This publication is part of the project "Self-induced spin glasses – a new state of matter) (OCENW.KLEIN.493) of the research programme KLEIN which is (partly) financed by the Dutch Research Council (NWO). B.V.

acknowledges funding from the Radboud Excellence fellowship from Radboud University in Nijmegen, the Netherlands. O.E. acknowledges support from the Swedish Research Council, the Foundation for Strategic Research (SSF) and the Swedish Energy Agency (Energimyndigheten). A.Be. and O.E. acknowledges the Knut and Alice Wallenberg Foundation and eSSENCE. A.Ba. acknowledges financial support from the Knut and Alice Wallenberg Foundation through Grant No. 2018.0060. The numerical simulations were enabled by resources provided by the Swedish National Infrastructure for Computing (SNIC). M.I.K and O.E. acknowledge the European Research Council via Synergy grant 854843 (FASTCORR).

**Author contributions**

B.V., L.Ni., and U.K. performed the experiments. D.W. and A.A.K. designed and participated in the experiments. B.V. and L.Ni. performed the experimental analysis. A.Ba. performed the complexity analysis. A.Be., D.I., L.Nor., M.I.K., and O.E. performed and participated in the theoretical calculations. B.V., D.W., A.Ba., A.Be., and A.A.K. wrote the manuscript, while all authors provided input to the manuscript.

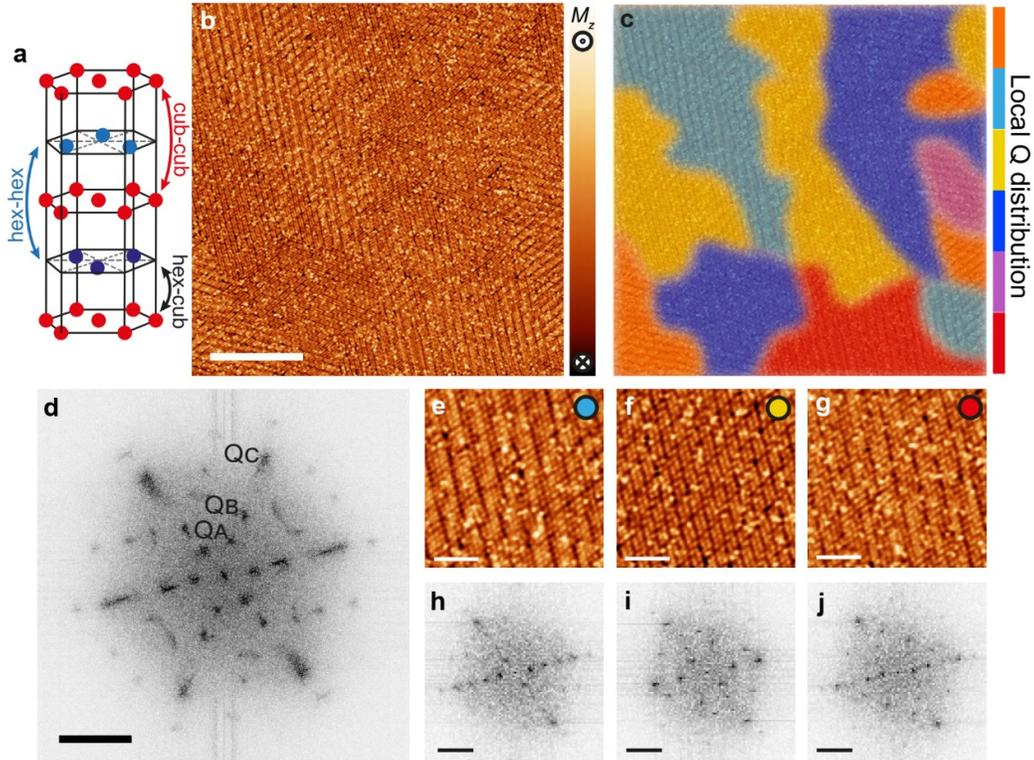

**Figure 1: Spin-Q glass state of Nd(0001) at 5 K.** a. dhcp structure of Nd(0001) characterized by magnetic interactions between both cubic (red) and hexagonal sublattices (blue) . b. Magnetization image based on SP-STM of Nd(0001) at $T$ = 5.2 K. ($I_t$ = 100 pA, scale bar: 50 nm). c. False-color schematic representing the different distributions of local Q order present in the magnetization image in b. d. Corresponding Q-space image (scale bar: 3 nm$^{-1}$) of (b) illustrating the various Q-pockets observed at $T$ = 5.2 K. e-j. Zoom-in real space/Q space images of b showing regions of local Q order defined by a particular set of Q vectors (e-g: scale bar: 10 nm, h-j: scale bar: 3 nm$^{-1}$). The colored dots correspond to the marked regions in c.

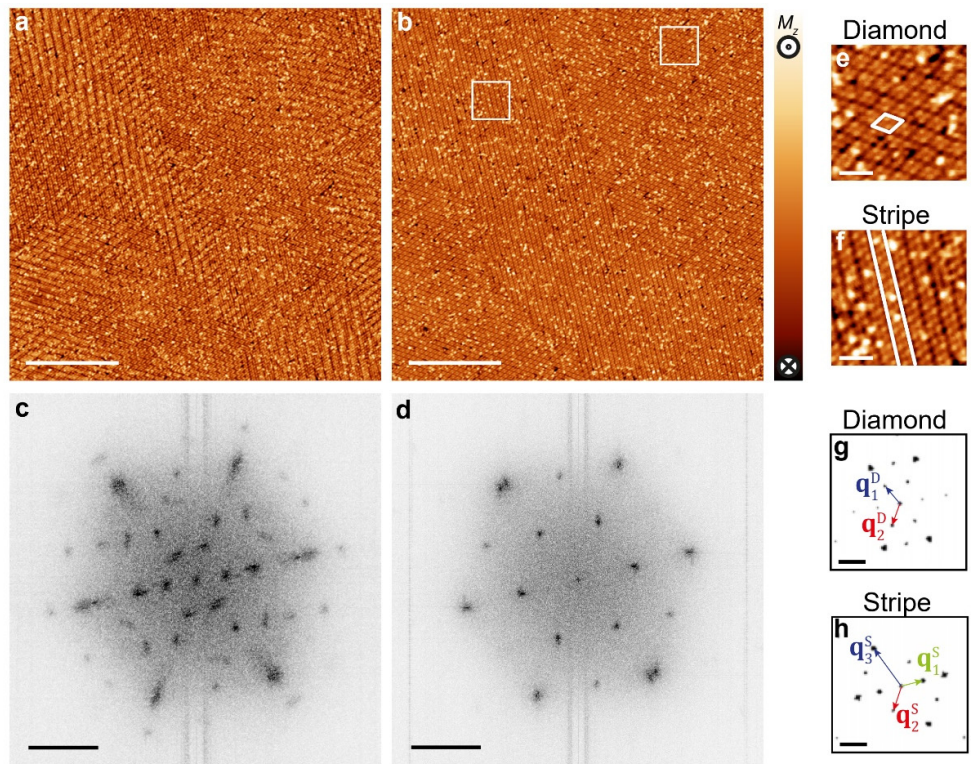

**Figure 2: Emergence of long-range multi-Q order from the spin-Q glass state at elevated temperature.** a,b. Magnetization images of the same region at $T$ = 5.1 K and 11 K, respectively ($I_t$ = 100 pA, a-b, scale bar: 50 nm). c,d. Corresponding Q-space images (scale bars: 3 nm$^{-1}$), illustrating the changes from strong local (i.e. lack of long-range) Q order toward multiple large-scale domains with well-defined long-range multi-Q order. e,f. Zoom-in images of the diamond-like (e) and stripe-like (f) patterns (scale bar: 5 nm). The locations of these images is shown by the white squares in b. g,h. Display of multi-Q state maps of the two apparent domains in the multi-Q ordered phase, where (g) corresponds to the diamond-like pattern and (h) corresponds to the stripe-like pattern (scale bar: 3 nm$^{-1}$). Those multi-Q state maps were extracted from the close-up views (40 x 40 nm²) and their FFT shown in Supplementary Fig. S4. A similar filtering procedure than in Supplementary Section S4 was performed.

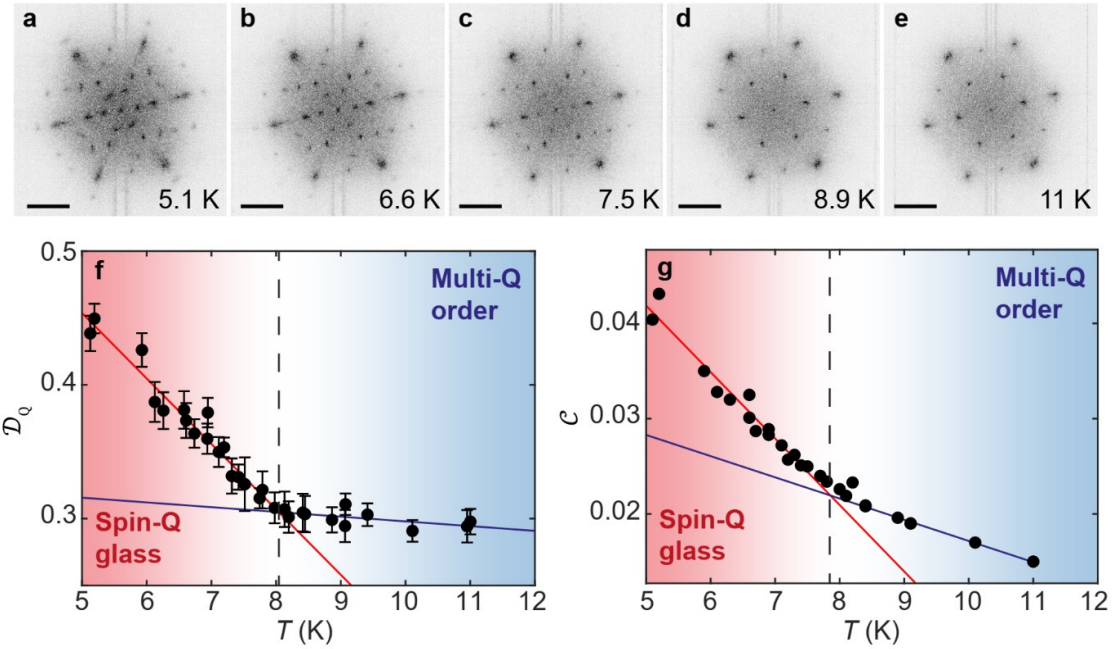

**Figure 3: Multi-Q magnetic transition and extracted $T_G$.** a-e. Q-space images of the same 200 x 200 nm² region at the indicated temperatures. (scale bar: 3nm⁻¹) f-g. Q-state divergence (f) and the value of the complexity (g) as a function of the temperature extracted from the real-space images used for (a-e). For definition of both properties, see text. For both plots, the red/blue lines correspond to linear fits in the spin-Q glass phase and multi-Q phase, respectively.

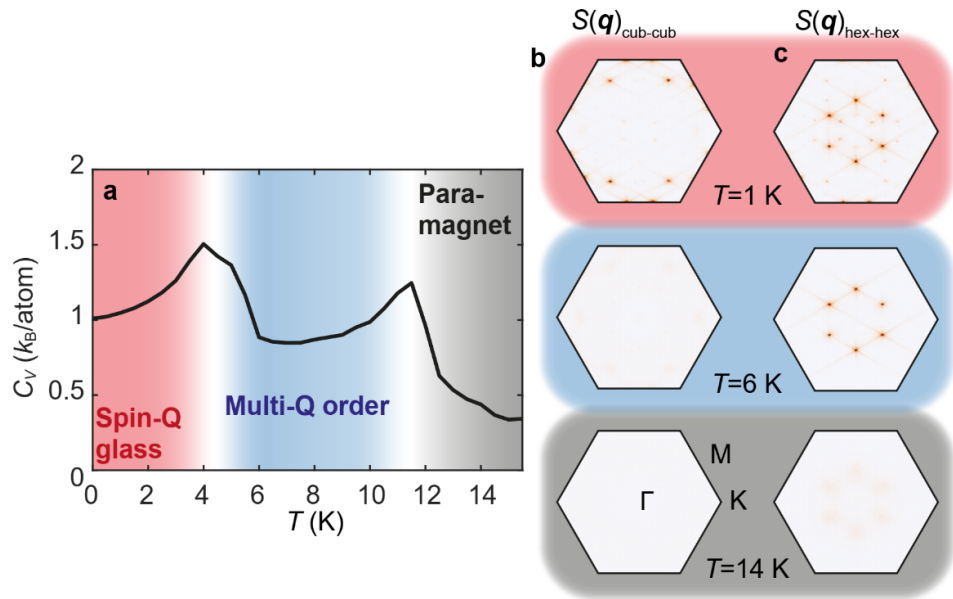

**Figure 4: Atomic spin dynamics simulations of the specific heat and sublattice-resolved S(Q).** a. Specific heat ($C_V$) as a function of temperature. Each peak corresponds to a phase transition at the indicated temperature. b. Sublattice resolved static structure factor ($S(q)$) for the (c) the cubic-cubic interactions and (d) the hexagonal-hexagonal interactions at the indicated temperatures selected for each phase. The BZ for all plots is indicated. (Darker contrast corresponds with higher values.)

# Supplementary information: Thermally-induced magnetic order from glassiness in elemental neodymium

Benjamin Verlhac[1], Lorena Niggli[1], Anders Bergman[2], Umut Kamber[1], Andrey Bagrov[2], Diana Iuşan[2], Lars Nordström[2], Mikhail I. Katsnelson[1], Daniel Wegner[1], Olle Eriksson[2,3], Alexander A. Khajetoorians[1,*]

**S1. Sample preparation**

Nd(0001) islands were epitaxially grown on a clean W(110) crystal as described in Ref.[1]. The W(110) substrate was cleaned by repeated cycles of annealing at $T$ = 1250°C in an oxygen atmosphere (p = 2 x $10^{-7}$ mbar) and flashing at $T$ = 2400°C. The Nd source material was purchased from AMES laboratory (www.ameslab.org; purity 4N), which was melted and then thoroughly degassed under UHV conditions inside the crucible of an electron beam evaporator, in order to further reduce the contamination from oxygen and hydrogen. For the growth of Nd films, the material was sublimated from the electron beam evaporator and deposited onto the W(110) kept at room temperature. Subsequent annealing at $T$ = 750°C for 10 minutes resulted in Stranski-Krastanov (SK) growth with an islands thickness of ~100 ML. Cr and W tips were chemically etched and field emission was performed on a clean W(110) surface to clean and shape the tip apex. For measurements with Nd-coated W tips, the W tip was coated by indentation into a Nd island.

**S2. Aging dynamics: 5 ↔ 10 K cycles**

We studied the response of the magnetic state of Nd when the temperature is repeatedly cycled between $T$ = 5 K and $T$ = 10 K. Fig. S1 shows the magnetization images (a,b) and their corresponding Q-space images (c,d) of the same area at $T$ = 5 K taken before and after warming up the sample to $T$ = 10 K. The area shown here is from a sample that presented a slightly higher defect density (0.0038 ML instead of 0.0025 ML). Nonetheless, the short-range order was still observable and a comparison of Fig. S1a with Fig. S1b leads to the same observations as the comparison of Fig. 1b with Fig. 2a, namely that there are distinct changes in the pattern distributions. Likewise, the comparison of the corresponding Q-space images (Fig. S1c and d) shows similar aging dynamics as already found from the comparison of Fig. 1d and Fig. 2c, marked by a redistribution of the spectral weight within the Q-pockets.

In Fig. S2, we show two iterations of magnetization images and their corresponding Q-space images at $T$ = 10 K before (a,c) and after (b,d) cooling down to $T$ = 5 K, respectively. The magnetization images look nearly identical with the only noticeable difference being the relocation of a domain wall on the top part of the image. No clear difference is present in the Q-space images. From this comparison, we conclude that no aging dynamics is observable at $T$ = 10 K.

To reproduce and quantify the observations, we performed an analysis of Q-space images from an entire series of temperature cycles using the Jensen-Shannon divergence[2,3]. Details on this divergence analysis are given in section S4. We took images from a sequence of seven 5-10 K temperature cycles and separated them in two sets, one being the sequence of $T$ = 5 K images, the other the sequence of $T$ = 10 K images. For each temperature set, we then used the Jensen-Shannon divergence $\mathcal{D}_{JS}(n, n+1)$, where $n$ is the image index, to quantify the differences between two successive Q-space images, respectively. Fig. S3 shows how the resulting $\mathcal{D}_{JS}(n, n+1)$ evolves as a function of $n$ at the two respective temperatures. The consistently lower divergence at $T$ = 10 K indicates the absence of changes in the spectral weight of the wave vectors that define the multi-Q ordered state. We notice that the value is not zero, due to the presence of background noise stemming from the STM measurement itself as well as defects on the surface. We also expect a small contribution due to domain wall motion. The higher divergence values at $T$ = 5 K quantify the observation that successive Q-space images are different from each other, reflecting aging dynamics which is marked by spectral weight redistribution of the Q-pockets.

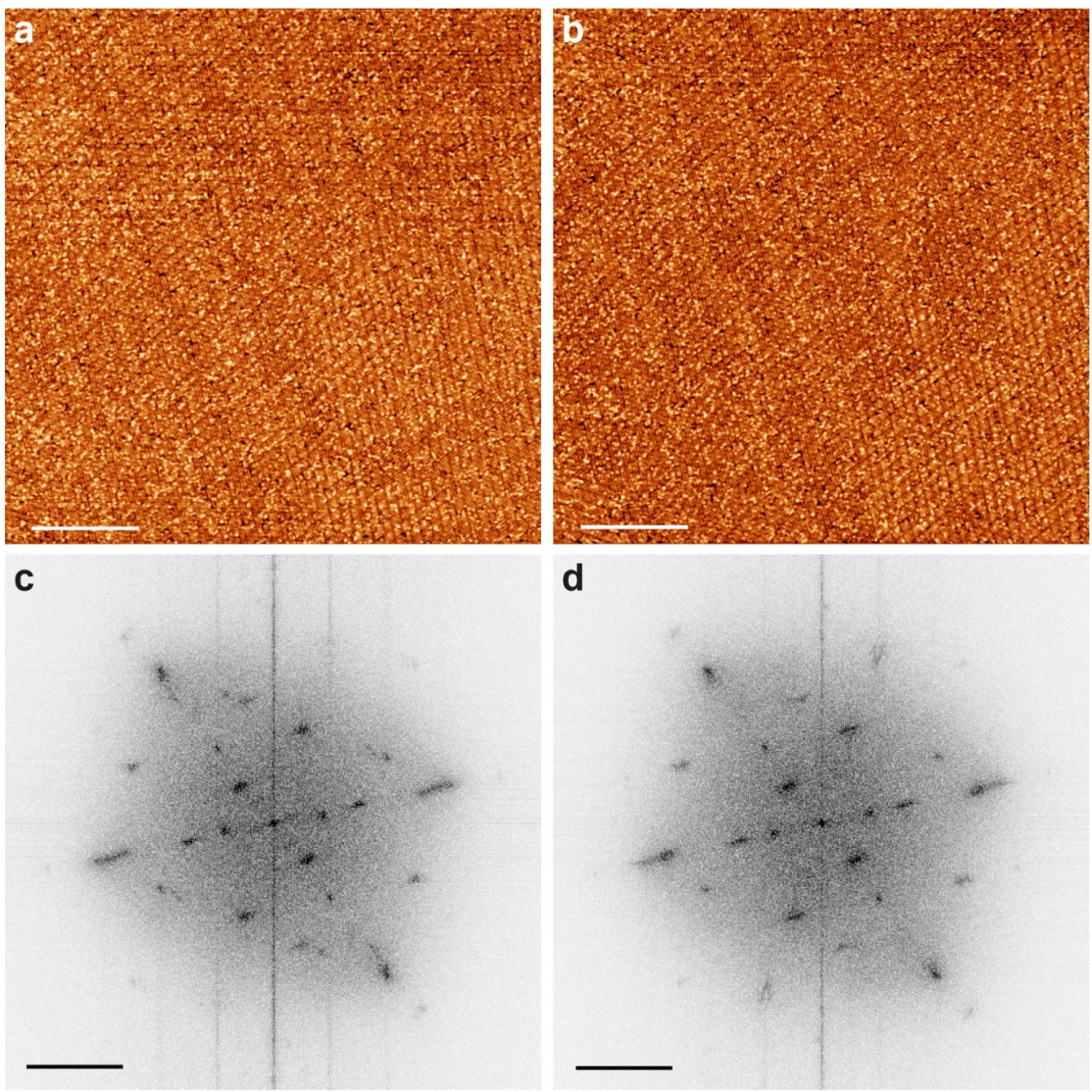

**Figure S1:** (a,b) Magnetization images of the same region at $T$ = 5.1 K before (a) and after (b) warming up the sample to $T$ = 10 K ($I_t$ = 200 pA, scale bars: 50 nm). (c,d) Corresponding Q-space images (scale bar: scale bars: 3 $nm^{-1}$).

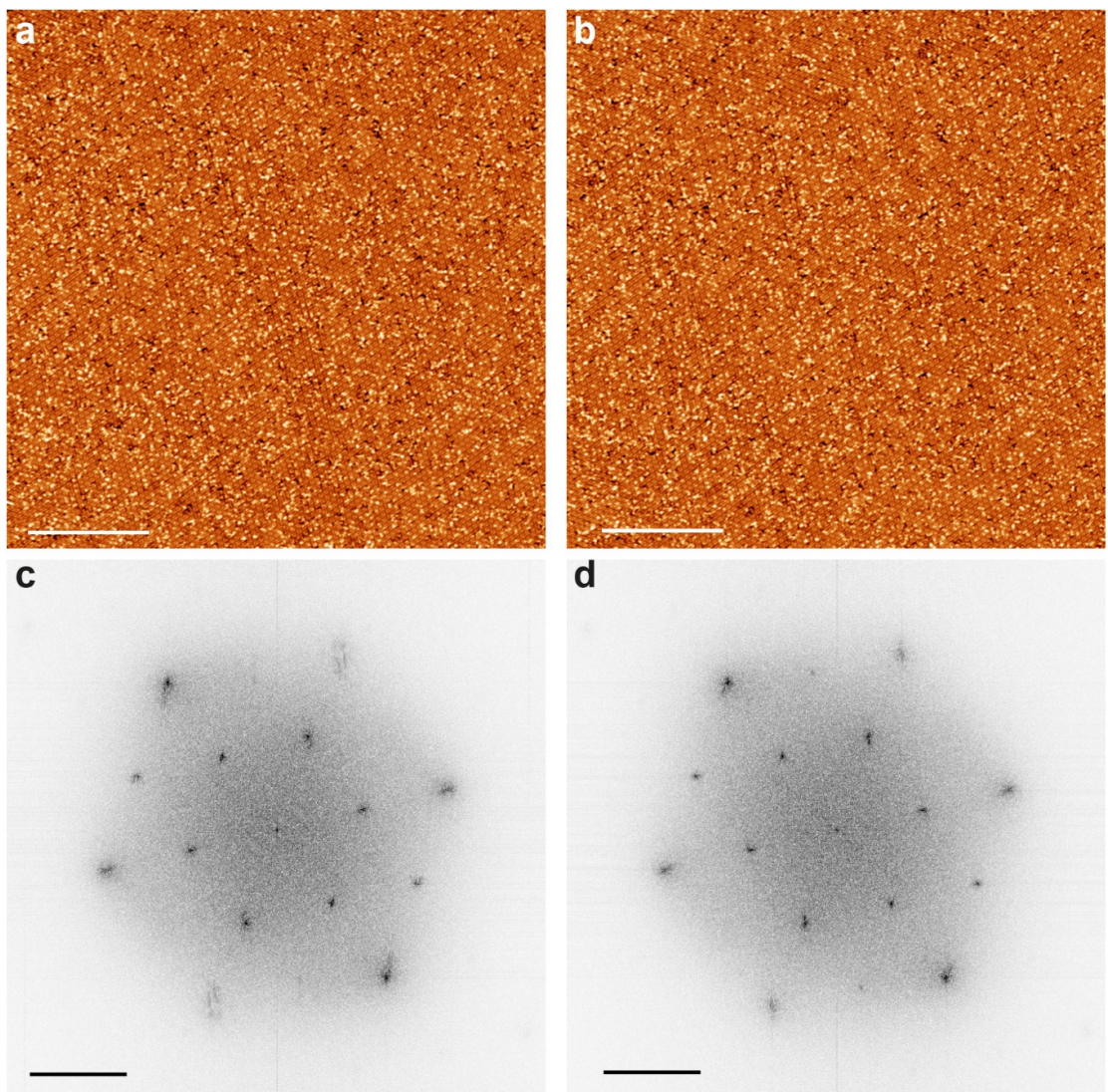

**Figure S2:** (a-b) Magnetization images and the corresponding Q-space images (c-d) of the same region at $T$ = 10 K ($I_t$ = 200 pA, a-b, scale bar: 50 nm, c-d, scale bar: 3 nm$^{-1}$)

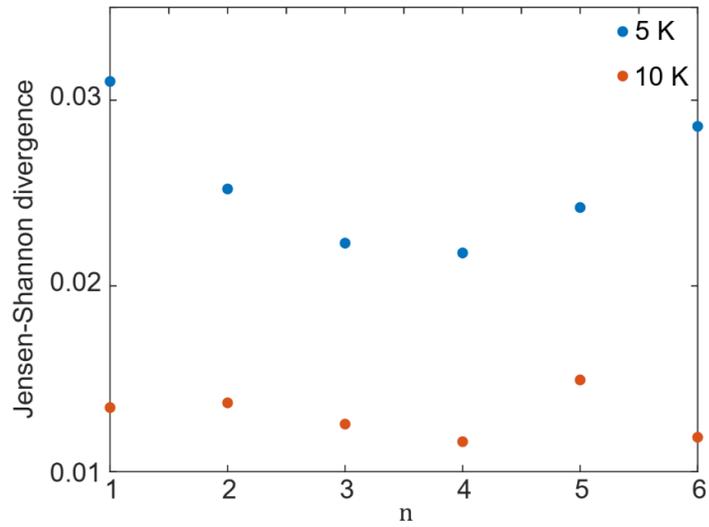

**Figure S3:** Evolution of the Jensen-Shannon divergence between two successive images at the same temperature. Blue: $T = 5$ K. Red: $T = 10$ K.

### S3. Additional magnetization images

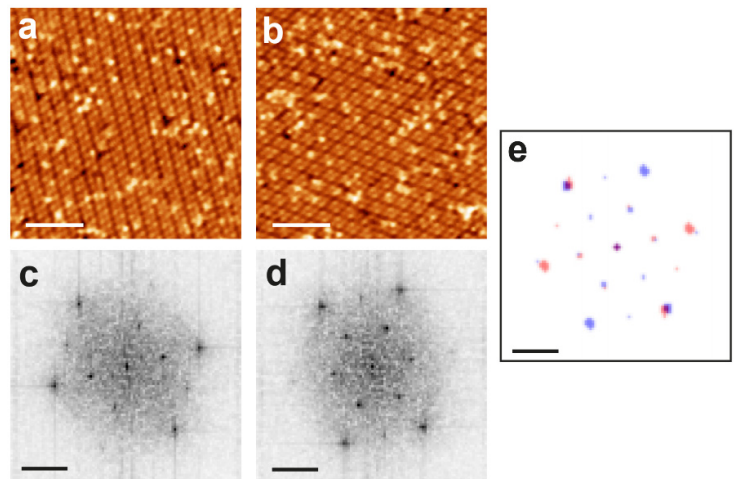

**Figure S4:** (a-b) Magnetization images and the corresponding Q-space images (c-d) of two different domains at $T = 11$ K ($I_t = 100$ pA, a-b, scale bar: 10 nm, c-d, scale bar: 3 nm$^{-1}$). e. Superposition of the Q-states of the two domains (Stripe phase in red and diamond phase in blue, scale bar: 3 nm$^{-1}$).

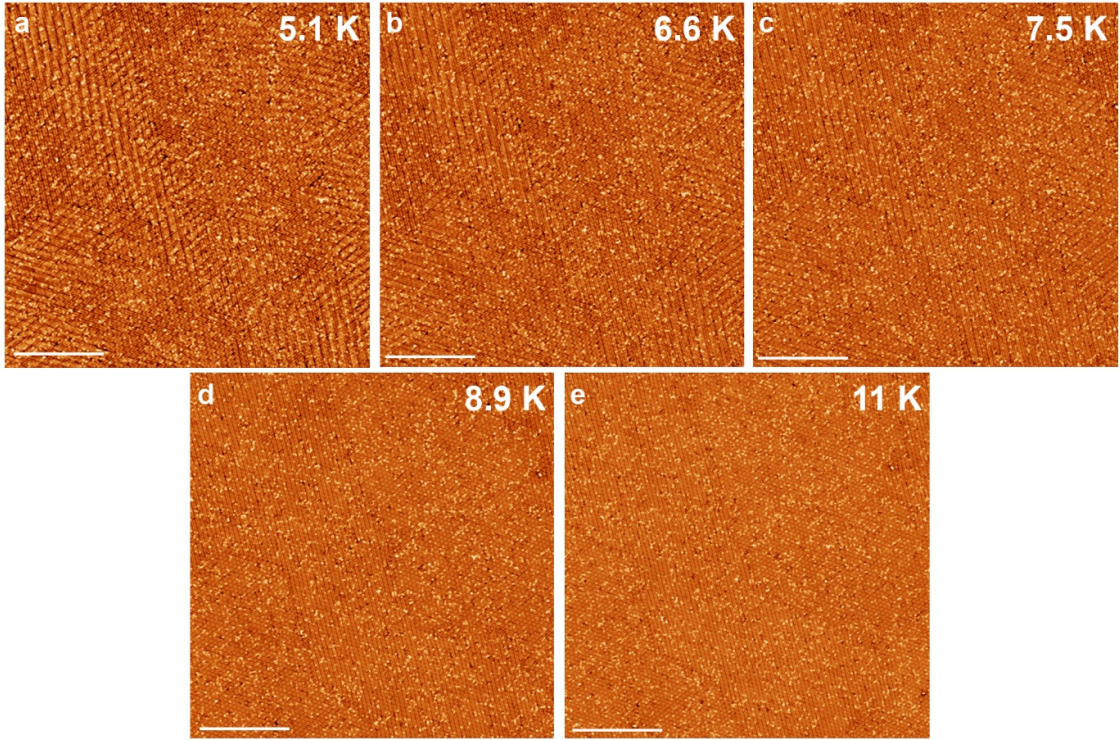

**Figure S5:** a-e. Magnetization images of the same region at the indicated temperatures (scale bar: 50nm).

**S4. Q-state divergence**

This section is devoted to describing the Q-state divergence analysis we performed to obtain a quantification of the phase transition. We detail the successive steps that allowed us to compute $\mathcal{D}_Q(T)$. Firstly, this analysis is based on the Jensen-Shannon divergence, $D_{JS}$, which is a mathematical tool quantifying the differences between two normalized discrete distributions, $P$ and $Q$:

$$\mathcal{D}_{JS}(P, Q) = \frac{1}{2} \sum_i \left( P_i \log\left(\frac{P_i}{M_i}\right) + Q_i \log\left(\frac{Q_i}{M_i}\right) \right)$$

with $M = \frac{P+Q}{2}$ the average distribution between $P$ and $Q$.

The Jensen-Shannon divergence is a well-established tool in information theory and possesses all the properties of a topological distance or metric[2,3], which procures an accurate measure of the differences between two distributions. An additional property of $D_{JS}$ is that it is bounded by 0 and 1 when a base-2 logarithm is used and these values are reached when $P$ and $Q$ are identical ($D_{JS}$ = 0) or two completely different distributions ($D_{JS}$ = 1). This property reveals itself to be practical in the interpretation of the Q-state divergence $D_Q(T)$ by looking at its proximity to the boundaries.

This analysis is based on extracting and comparing *local* Q-state distributions from the measured magnetization images. Here we describe the method that we used to acquire such *local* Q-state distributions from close-up views of a large magnetization image. We first select a region of 22 x 22 nm² from any magnetization image (example shown in Fig. S6a at $T$ = 11 K). We compute its FFT in a 15 x 15 nm$^{-2}$ window (Fig. S6b) and then apply a sequence of filters. We start by applying a gaussian filter ($\sigma$ = 0.6) to reduce the noise level. The next step is to apply a threshold filter that allows to identify the Q-spots from the FFT via binarization. However, as the background noise is not constant but becomes larger toward the center of the FFT, we used a threshold filter that discriminates the signal from this background by adapting the threshold. The algorithm related to this filter sets a different threshold parameter for each pixel by calculating statistics of its neighboring pixels (ref.[4] for more info). The sensitivity parameter of this filter is set to S = 0.16. The resulting Q-state map is then blurred with another gaussian filter ($\sigma$ = 0.5) in order to smooth the edges of the extracted Q-spots. The result is shown in Fig. S6c. This procedure allows to capture most of the relevant features but is limited in a sense that one need to optimize the noise reduction and the threshold sensitivity to avoid as much as possible getting spots due to remaining noise. However, the filtering procedure is not perfect as we can see the presence of spots that come from noise lines present in the FFT (Fig. S6c). Another optimization we made concerns the size of the real space regions one can extract from the magnetization images. We chose 22 by 22 nm² to be the smallest region size so that we ensure that we obtain the most correct description of the local order (especially in the Q-glass phase) while having enough pixel density in the FFT to resolve the Q-spots.

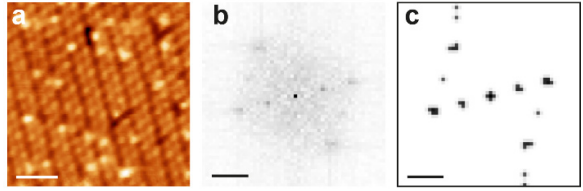

**Figure S6:** a. Close-up view of a magnetization image at $T$ = 11 K ($I_t$=100pA, scale bar = 5 nm). b. Corresponding Q-space image (scale bar = 3 nm$^{-1}$). c. Q-state map resulting from the sequence of filters applied on the Q-space image (scale bar = 3 nm$^{-1}$).

In order to determine the Q-state divergence within a large magnetization image at a given temperature, we start by selecting 40 random 22 by 22 nm² regions, as exemplified in Fig. S7a. The selection was done by defining a grid of 50 x 50 positions and selecting randomly the x and y positions of the center of each region. This leads to 2500 possibilities of positions of the regions. To ensure that there is not too much overlap and that the regions are more spread over the magnetization image, we excluded double values for x and y. From each of these regions, we extracted the corresponding Q-state maps with the method described above. We then calculated the Jensen-Shannon divergence of the local Q-state map of the region indexed $n$ from the average Q-state map of the regions 1 to $n-1$:

$$D_n = \mathcal{D}_{JS}(Q_n, \langle Q_i \rangle_{n-1})$$

where $\langle Q_i \rangle_{n-1} = \frac{1}{n-1}\sum_{i=1}^{n-1} Q_i$. This step provides then a sequence $D_n$ which is our first sample of statistics on how different a local Q-state can be to the average Q-state of the scanned area.

The second step of this statistical process consists of iterating this random selection. By iterating 30 times the selection of 40 random boxes, 30 different sequences $D_n$ are obtained (plotted as shades of blue and green in Fig. S7b and Fig. S7c) that we averaged to obtain the red set of points. As can be seen, this red dataset converges rapidly with progressing $n$ to a value that corresponds to the average divergence between the Q-state map of a small region and the average Q-state map of a magnetization image at a given temperature. In order to acquire a single mean value of the Q-divergence for a particular magnetization image at a given temperature, $D_Q(T)$, we averaged the red dataset from index $n$ = 15 to 40, where the standard deviation is used as the error bar in Fig. 3f. Fig. S7d and Fig. S7e show the average Q-state maps of all the 1200 boxes selected from a magnetization image taken at $T$ = 5 K and $T$ = 11 K, respectively. We notice that these maps reproduce nicely the Q-spots observed in the Q-space images of the whole magnetization images, which are shown in Fig. 2, validating the threshold filtering and statistical methods that were used.

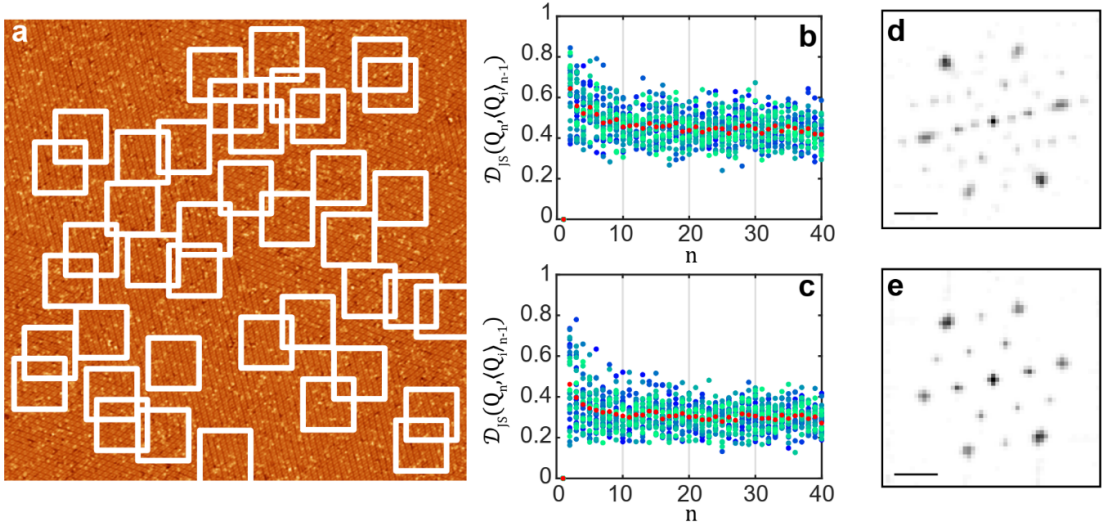

**Figure S7:** a. Magnetization image of the multi-Q order at $T$ = 11 K with rectangles representing the cropped 22 x 22 nm² regions selected randomly during an iteration of the Q-state divergence analysis. b. Result of the calculation of $\mathcal{D}_{JS}(Q_n, \langle Q_i \rangle_{n-1})$ over 30 iterations of the random selection of 40 regions (each iteration corresponds to a shade of blue and green) realized from a magnetization image acquired at $T$ = 5 K. The red dataset corresponds to the average over all the iterations. c. Same plot but for a magnetization image acquired at $T$ = 11 K. d,e. Average of all the 30 times 40 Q-state maps used for the analysis of the $T$ = 5 K (d) and $T$ = 11 K (e) magnetization image, respectively.

**S5. Complexity Analysis**

In this section, we describe the methods used to identify the transition point between the spin Q-glass state and the ordered multi-Q state directly from the real-space data. In Ref. [5], a measure of structural complexity of patterns has been suggested. This measure is based on counting features corresponding to different spatial scales and, among other things, has been shown to be an effective tool to detect phase transitions in physical systems without any prior knowledge of the transition nature. Referring the reader to the original paper for the detailed discussion, here we provide practical aspects of this procedure. The original image $P_0$ (the magnetization image in this context) undergoes a step-by-step renormalization group procedure $P_0 \to P_1 \to P_2 \to \cdots$, and the stack of coarse-grained images $\{P_k\}$ is retained. Patterns in images can be interpreted as contrast differences occurring at certain scales. The coarse-graining procedure filters out the different patterns present in the image, from the smallest patterns to the largest as $k$ increases. We then quantify the "loss" of patterns between the step/scale $k$ and $k + 1$ by computing the deviation between $P_k$ and $P_{k+1}$ as follows:

$$C_k = \frac{1}{2} \int_D (P_{k+1} - P_k)^2 dx$$

where the integration goes over spatial domain $D$ which corresponds to the sum of all pixels in the bitmap STM magnetization image. $C_k$ is then called partial complexity, while $C = \sum_k C_k$ is the overall structural complexity. We computed the structural complexity of magnetization images at different temperatures using two different coarse-graining schemes. To be able to perform this procedure, we first used the image processing tool ImageMagick (https://imagemagick.org/index.php) to slightly upscale the magnetization images to reach a 2048 x 2048 pixel resolution (the original resolution is about 2000 x 2000, so this transformation does not induce many artifacts), and then performed 7 steps of coarse-graining. The domain $D$ corresponds then to 2048 x 2048 pixels and is the same for all the procedure. The first scheme is the same as the one used in Ref. [5] and based on simple averaging. At step $k = 1$, $P_0$ is divided into groups of 2 x 2 pixels and within each group the value at each pixel is substituted by the average pixel value of the group. The resulting image corresponds then to $P_1$. At step $k$, to construct the $P_k$ image, $P_{k-1}$ is divided into groups of $2^k$ x $2^k$ pixels, and their values are substituted with the average value within the group (note that the total size of the image remains the same over the course of the procedure; in the terms used above, domain $D$ is unchanged). Computing the corresponding overall complexity for each image in the dataset, we obtained the temperature-dependent complexity $C(T)$ that is shown in Fig. 3g.

To verify robustness of the result, we performed the same procedure, but with another coarse-graining scheme. In the magnetization images, disorder from defects appears as brighter pixels, and by letting darker signal dominate at coarser scales, we could eliminate, to a large extent, this defect contribution to the complexity computation. To do that, at step $k$, instead of averaging values of pixels in the $2^k \times 2^k$ groups, we can pick the darkest one of each group, and assign its value to all the pixels of the group. Fig. S8 provides an example of such a sequence. In Fig. S9, we show the comparison of the complexity for the two coarse-graining schemes. Both methods highlight the presence of two phases, with a clearer difference in the averaging coarse-graining scheme. We note that from these two methods we found comparable values of $T_G$ with the average coarse-graining scheme giving $T_G$ = 7.9±0.2 K, while the dark spot coarse-graining scheme gives $T_G$ = 8.1±0.2 K.

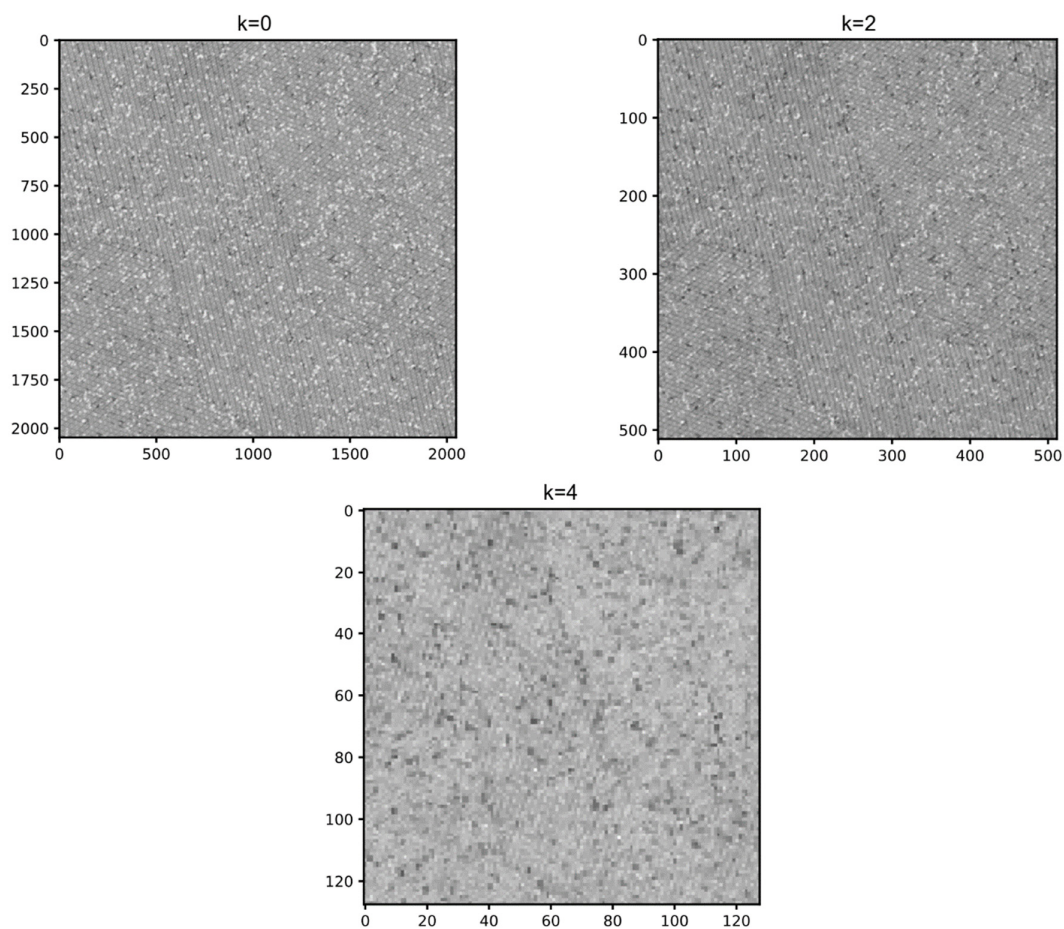

**Figure S8:** Coarse-grained images of the same 200 x 200 nm² area at $T$ = 11 K for $k$ = 0, 2 and 4 using the dark spot coarse-graining scheme.

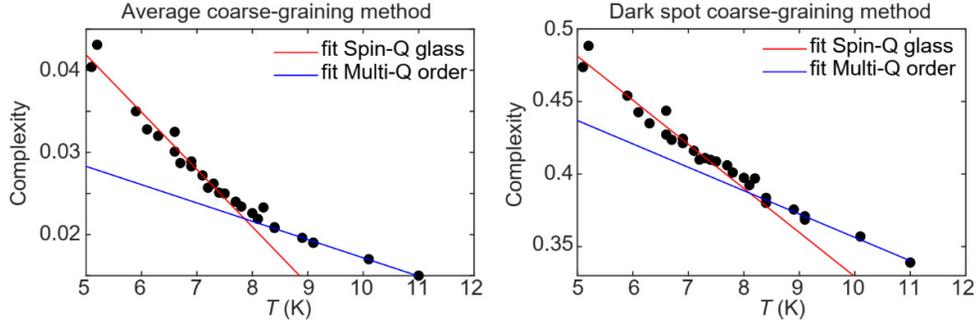

**Figure S9:** Plot of the complexity as a function of the temperature for the two coarse-graining schemes.

### S6. Two-time autocorrelation function:

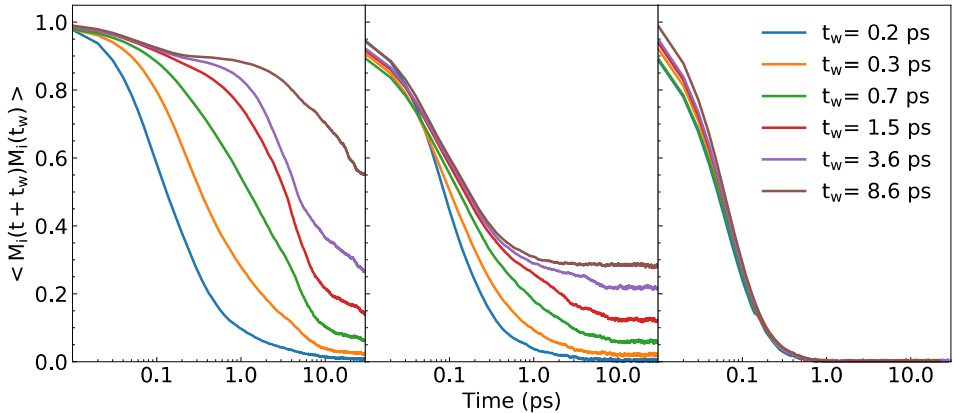

**Figure S10:** Two-time autocorrelation functions for $T$ = 1, 6 and 14 K.

In order to identify the presence (or lack) of glassy dynamics in the magnetic phases of Nd, we studied the two-time autocorrelation function $C(t_w, t) = \langle m_i(t + t_w) \cdot m_i(t_w) \rangle$, which depends on the simulated time $t$ and the waiting time $t_w$. In practice the simulations start from a random distribution of atomic spins, that are allowed to relax according to the atomistic Landau-Lifshitz-Gilbert equation for a time $t_w$. After that the two-time autocorrelation was calculated for a time period $t$. The autocorrelation function was simulated using ASD simulations following the same protocol as in Ref. [1], and the method is partly described in the Method section of the main text as well as in the supplementary information of Ref. [1]. The results of the autocorrelation simulations are shown in Fig. S10 where the results for temperatures $T$ = 1 K, 6 K, and 14 K are presented. At $T$ = 1 K the autocorrelation functions show multiple relaxation times depending on the waiting time $t_w$, which is a typical signal of spin-glass behaviour. In contrast, at $T$ = 6 K, the relaxation behavior is similar regardless of the waiting time. The magnetic state at $T$ = 6 K is identified to be an ordered multi-Q spin spiral state. The relaxation behavior presented in the middle panel of Fig. S10, show that curves from different waiting times stabilize at different values, something that can be explained by the fact that the finite-size simulations performed here have no anisotropy or other effects that keep the system from performing global rotations due to the thermal fluctuations. At $T$

= 14 K, the system is paramagnetic and the rapid relaxation rate comes from the thermally induced disorder in the simulated system.